\documentclass[aps,prl,groupedaddress]{revtex4}

\begin{document}

\title{Finiteness of the universe and computation beyond Turing computability}

\author{Tien D. Kieu}

\affiliation{CAOUS, Swinburne University of Technology, Hawthorn 3122, Australia}
\email{kieu@swin.edu.au}
\date{\today}

\begin{abstract}
We clarify the confusion, misunderstanding and misconception that the physical finiteness of
the universe, if the universe is indeed finite, would rule out all hypercomputation, the
kind of computation that exceeds the Turing computability, while maintaining 
and defending the
validity of Turing computation and the Church-Turing thesis.
\end{abstract}
\maketitle

Through private communication with some individuals, we have encountered some confusion,
misunderstanding and misconception that the physical finiteness of the universe, if the
universe is indeed finite, would rule out all hypercomputation, the kind of
computation that exceeds the Turing computability.  And now this misleading
thinking has somehow made its way to formal presentation in~\cite{S}.  We
would like to take this as an
opportunity to publicly present our arguments, for the record, against such misconception.
For that purpose, we pose
below three questions and then give our answer to each one.

\section*{Is the universe finite?}
We do not know for sure, even though it would not surprise us if the universe is finite.

This is an important physics question and will surely be investigated and debated
thoroughly in the years to come.  However, 
in encountering this finiteness presumption or any
{\em (yet to be confirmed) model} of quantum measurement which
implies such finiteness in a discussion of hypercomputation,
one should keep it in mind that this is only an assumption or an unconfirmed model,
and not a fact.

Let us recall that we have
explicitly assumed that the universe is infinite in discussing our quantum algorithm
for Hilbert's tenth problem~\cite{contphys}.  This assumption is only for the
convenience in presenting our algorithm, so that we could avoid 
the need of introducing unnecessary
distractions.  However, we have also stated elsewhere in the same paper that
it is sufficient to have the dimensions of the underlying Hilbert space finite
but unbounded.

\section*{Would such finiteness maintain the status quo of the Church-Turing
thesis?}
No.

This is clearly seen by taking the arguments of~\cite{S} which lead to
the result that Chaitin's $\Omega$~\cite{Chaitin} (see also~\cite{FI}) 
is not computable because a
physically finite universe would allow us to physically compute
(by some unspecified means)
only a finite number of binary digits of the number (once a programming
language has been specified).  Such arguments are of course correct but,
unfortunately, are also applicable to
more `normal' and `ordinary' numbers such as $\pi$ or $e$: with finite physical
resources, any Turing machine can physically compute only some finite number
of binary digits of any real number!  In this way, we would have to conclude
that $\pi$, for example, is {\em noncomputable} too!  Also, `most' rational
numbers would have been classified noncomputable!
Clearly, this is too
restrictive and not very useful a discussion of computable numbers.  In fact,
with such restriction, one would not need the concept of effective computation,
of recursive functions in general.  And
neither one would need the thesis of Church Turing at all--let alone hoping
that the physical finiteness of the universe would support the
thesis itself as wishfully presented in~\cite{S}.  After all, with finite physical
resources one can physically represent, in binary form say, only some large but finite
number/integer, whether it is in Turing computation or hypercomputation.
Full stop.   For any number larger than this physical limit, only abstract
mathematical representations can exist.

The point we want to draw attention to here is that such use of {\em physical}
finiteness of the universe is not in the spirit of
even {\em mathematical} Turing computation--let alone
hypercomputation--and not at all fruitful in the context of
{\em mathematical} computability.

This leads us to a more useful and relevant question next.

\section*{Would such finiteness render all hypercomputation ineffective?}
No, in as much as physical finiteness would not render Turing computation
ineffective.

Recall that Turing machines are abstract constructs in which finite but
unbounded tapes are required for the operation.  The tapes can be lengthened
as much as necessary during the computation.
Parallely similar to the lengths of these tapes in Turing machines are the dimensions of
the underlying Hilbert spaces in our quantum adiabatic algorithm for Hilbert's tenth
problem~\cite{Kieu}.  Given any Diophantine equation, the algorithm looks for the
global minimum of the square of the Diophantine polynomial (since knowing this
minimum, we can then decide if the equation has a non-negative integer 
solution--i.e. when and only when this global minimum is zero).  It is easily seen that
the global minimum for the square of any given Diophantine polynomial has to 
take place at some {\em finite} values for the polynomial variables. 
This fact is also reflected in the {\em finite} energy of the ground state to be obtained
in our quantum adiabatic algorithm.
As a result, the dimension of the underlying Hilbert space
need be only finite (but sufficiently large).  We demonstrate in~\cite{KieuOrlando}
how to find such sufficiently large dimesions.

The physical finiteness of the universe would of course impose some upper limit
on the number of dimensions one can physically realise.  But as we know when
in a Turing computation the end of a Turing tape has been reached and cannot 
be lengthened further due to lack of resources, we would also know when the
upper dimensions of the computation Hilbert space have been physically
arrived at.  At that point, the computation has to be abandoned before we can
obtain the final result.  At no time, however, the physical
finiteness of the universe should lead us to the wrong computation result; it simply
would not allow us to complete the computation for some group of Diophantine
equations.

In summary, the physical finiteness of the universe should not impose any limitations 
on hypercomputation more than those which it would already impose on Turing computation
since, in the end, {\em all} computation is physical.  Because of this
indiscrimination, it is logically inconsistent and wrong to use
the finiteness arguments to rule out hypercomputation while still maintaining 
and defending the validity of Turing computation and the Church-Turing thesis.
On the other hand, the probable physical finiteness should not and cannot 
stop us from investigating hypercomputation as it has not deterred us from 
studying Turing computation (or mathematics in general).

\vspace{1in}

\noindent
{\bf Notes added on 16 April 2004:}\\
After my posting of this paper, Prof Srikanth has sent me a revised
version of {\tt quant-ph/0402128} and pointed out to me that his work
concerned with a different kind of finiteness which limits the number of 
terms allowed in a quantum coherent linear superposition.  This finiteness 
of quantum paralellism, however, is non-standard and constitutes another 
postulate different from the usual
von Neumann postulate of measurement in quantum mechanics.
\end{document}